\documentclass[a4paper,10pt]{article}
\pdfoutput=1 % if your are submitting a pdflatex (i.e. if you have
\usepackage{jcappub}
\usepackage{mathrsfs}
\usepackage{amsmath,mathtools}

\usepackage{graphicx,multirow,subfigure}

\newcommand{\mat}[1]{\begin{pmatrix} #1 \end{pmatrix}}
\newcommand{\n}{\nonumber\\}

\def\lsim{\raise0.3ex\hbox{$\;<$\kern-0.75em\raise-1.1ex\hbox{$\sim\;$}}}
\def\gsim{\raise0.3ex\hbox{$\;>$\kern-0.75em\raise-1.1ex\hbox{$\sim\;$}}}

\usepackage{cancel}
\usepackage[utf8]{inputenc}

\date{Compiled on \today~ at \currenttime}
\def\be{\begin{equation}}
\def\ee{\end{equation}}
\def\bea{\begin{eqnarray}}
\def\eea{\end{eqnarray}}

\usepackage{pstricks}
\usepackage{epsfig}
\title{Bound States of Right-handed Sneutrino Dark Matter }
\author[a,b]{A.~Elsheshtawy}
\author[b]{and S.~Khalil}

\affiliation[a]{\small Department of Mathematics,Faculty of Science, Ain Shams University, Cairo 11566, Egypt.}
\affiliation[b]{\small Center for Fundamental Physics, Zewail City of Science and Technology, 6 October City, Giza 12578, Egypt.}

\abstract{
The relic abundance of the lightest right-handed sneutrino dark matter in supersymmetric  B-L extension of the standard model is revisited. We argue that it can form a bound state through the exchange of light $B-L$ Higgs boson and a large non-perturbative Sommerfeld effect is obtained. We emphasized that with this effect the relic abundance of right-handed sneutrino will lie  within the observational limits and the right-handed sneutrino will remain a viable dark matter candidate of mass order up to 1.5 TeV. 

}

\begin{document}
\maketitle
%%%%%%%%%%%%%%%%%%%%%%%%%%%%%%%%%%%%%%%%%%%%%%%%%%%%%
\section{Introduction}

Dark Matter (DM) is one of the most important evidences for physics Beyond the Standard Model (BSM) of particle physics.  The relic abundance of the DM candidacy, since its decoupling time, is an essential issue for its validity to account for the DM observation. The latest results from the full-mission of Planck satellite indicated that the DM relic abundance is give by \cite{Aghanim:2018eyx}
\be
\Omega_{\rm DM} h^2 = 0.120 \pm 0.001.
\label{planck}
\ee
This tiny region imposed stringent constraints on the permitted parameter space of any BSM provides a DM candidate. Since the thermal relic abundance of DM is inversely proportional to its annihilation cross section, which is usually quite small due to the fact that most of new particles in BSM models are quite heavy to satisfy collider negative searches, the predicted relic abundance is much larger than the observed limits in Eq. (\ref{planck}).

On the other hand, a bound state of DM can be formed if it couples to a light force mediator. In this case, a non-perturbative effect, called Sommerfeld Effect (SE) would play an important role in enhancing the annihilation process, and thus reducing the relic density down to the target range of Planck measurements.  The formation of DM bound state through a scalar field, $\phi$, mediator requires the screening length, $1/m_\phi$, to be much larger than the Bohr radius of bound state , $\sim 2/\alpha\, m_{\rm DM}$,  which is the most probable distance between the right-handed sneutrinos  in the bound state. Thus, one gets the following conditions \cite{PhysRevA.1.1577, Braaten:2018xuw, Kang:2016wqi}:
\be 
\alpha\, m_{\rm DM} /m_\phi \gsim 1.64, 
\ee
where the dimensionless coupling $\alpha$ is given by $\alpha\equiv Y^2/16 \pi m_{\rm DM}^2$ \cite{Petraki:2015hla}, with $Y$ is the dimensionfull trilinear coupling between $\phi$-DM-DM. Therefore, this scenario entails: $m_{\rm DM} \gg m_\phi$. 
 
In this paper, we show that the lightest right-handed sneutrino, which is a natural candidate of DM in the B-L extension of the minimal supersymmetric standard model (MSSM), can form a bound state through the Yukawa potential produced by the mediation of the lightest Higgs boson, $h'$, which can be as light as 28 GeV, as shown in Ref.\cite{Cici:2019zir}.  The B-L Supersymmetric Standard Model (BLSSM) is based on the gauge group $SU(3)_C \times SU(2)_L \times U(1)_Y \times U(1)_{B-L}$ \cite{ Khalil:2006yi, Khalil:2007dr, Emam:2007dy, Basso:2008iv, Basso:2009gg, Perez:2009mu, Basso:2010as, Basso:2010yz, Khalil:2012gs, Khalil:2013in, DelleRose:2017ukx}.  This extension requires three right-handed neutrino superfields be added to cancel the $U(1)_{B-L}$ anomalies and to implement seesaw mechanism, which provides an elegant solution for the existence and smallness of the light neutrino masses \cite{Khalil:2006yi, Abbas:2007ag}. It has been shown that in this class of models, the lightest right-handed sneutrino can be the lightest supersymmetric particle (LSP), thus a viable candidate of Dark Matter (DM)                    \cite{DelleRose:2017uas,  Khalil:2011tb}. In addition, the CP-even Higgs boson associated with breaking $U(1)_{B-L}$ symmetry can be quite light \cite{ Hammad:2016trm,Abdallah:2014fra}. This Higgs boson has a large coupling to the right-handed sneutrino, as they belong to the same sector of $B-L$ symmetry. Therefore, it is a potential candidate for mediating a bound state between the lightest right-handed sneutrinos. 

The paper is organized as follows. In section 2 we emphasize that the standard calculation of sneutrino relic abundance leads to results that exceed the observational limits. In section 3 we study the bound state of sneutrino through $h'$ exchange and the associated SE. In section 4 we calculate the relic abundance with the predicted SE and provide a benchmark points where the resulting $\Omega h^2$ is compatible with the stringent observational constraints for a large range of right-handed sneutrino mass. Our conclusions and final remarks are given in section 5. 

%%%%%%%%%%%%%%%%%%%%%%%%%%%%%%%%%%%%%%%%
\section{Right-handed Sneutrino Dark Matter}
The BLSSM model consists of the MSSM particle content and two chiral SM-singlet Higgs superfields ($\hat{\chi}_{1,2}$ with $B-L$
charges $Y_{B-L}=\mp2$, respectively), whose Vacuum Expectation Values (VEVs)
of their scalar components, $v'_1=\langle \chi_1\rangle$ and $v'_2=\langle \chi_2\rangle$, spontaneously break the $U(1)_{B-L}$, in addition to three
right-handed neutrino superfields ($\hat{N}_i$). The superpotential of the BLSSM is given by
\bea {W} &=&
(Y_u)_{ij}\hat{Q}_i\hat{H}_2 \hat{U}^c_j+(Y_d)_{ij}\hat{Q}_i\hat{H}_1\hat{D}^c_j+(Y_e)_{ij}\hat{L}_i\hat{H}_1\hat{E}^c_j
+ (Y_{\nu})_{ij}\hat{L}_i\hat{H}_2\hat{N}^c_j  + (Y_N)_{ij}\hat{N}^c_i\hat{\chi}_1\hat{N}^c_j\nonumber\\
&+& \mu \ \! \hat{H}_1\hat{H}_2+\mu' \ \! \hat{\chi}_1\hat{\chi}_2.
\label{super-potential-b-l}
\eea%
The $B-L$  charges of the above superfields, the corresponding soft SUSY breaking terms, and the details of $B-L$ radiative symmetry breaking at TeV scale ($v' = \sqrt{{v'_1}^2 +{v'_2}^2} \gsim 7 $ TeV) can be found in Ref.~\cite{Khalil:2007dr}. The detail of this model and its phenomenological implications can be found in Ref.\cite{OLeary:2011vlq}.

The sneutrino mass matrix, in the basis ($\tilde{\nu}_L,\tilde{\nu}_L^\ast,\tilde{\nu}_R,\tilde{\nu}_R^\ast$), is approximately given by a $2\times 2$ block diagonal matrix, where the element $11$ of this matrix is given by the diagonal LH sneutrino mass matrix and the element $22$ represents the RH sneutrino mass matrix, $M_{RR}$, which is defined as 
\begin{eqnarray}
		M^2_{RR} & = & \mat{M_N^2 + m_{\tilde{N}}^2 + m_D^2+ \frac{1}{2}M_{Z'}^2\cos 2\beta' & M_N(A_N-\mu'\cot\beta')\\
		M_N(A_N-\mu'\cot\beta') & M_N^2+m_{\tilde{N}}^2+m_D^2+\frac{1}{2}M_{Z'}^2\cos 2\beta'}.
	\end{eqnarray}
where $M_N$ is the right-handed neutrino mass, which is proportional to the $B-L$ symmetry breaking VEV, {\it i.e.}, $M_N= Y_N v'_1 \sim {\cal O}(1)$~TeV, and $m_D=Y_\nu \langle H_2\rangle = Y_\nu v_2$, with $Y_\nu \lsim {\cal O}(10^{-6})$, to fulfill the smallness of  light neutrino masses~\cite{Abbas:2007ag}. The soft SUSY breaking parameters $m_{\tilde{N},\tilde{L}}$ and $A_{\nu,N}$ are the sneutrino, slepton scalar masses and trilinear couplings, respectively, which are given by universal values
at the Grand Unification Theory (GUT) scale and are determined at any scale by their Renormalization Group Equations (RGEs). Finally $\tan \beta'$ is defined as the ratio between the two $B-L$ VEVs, $\tan\beta'=v'_1/v'_2$, in analogy to the MSSM VEVs $(\tan\beta = v_2 /v_1)$. 

It is worth noting that the mixing between the right-handed sneutrinos and right-handed anti-sneutrinos  is quite large, since it is given in terms of $Y_N \sim {\cal O}(1)$. Thus, the eigenvalues of the right-handed sneutrino squared-mass matrix $M^2_{RR}$ are given by
\begin{equation}
m^2_{\tilde{\nu}_{\mp}} = M_N^2+m_{\tilde{N}}^2+m_D^2+\frac{1}{2}M_{Z'}^2\cos 2\beta'  \mp  \Delta m_{\tilde{\nu}_R}^2,
\label{eq:mass_splitting}
\end{equation}
where $\Delta m_{\tilde{\nu}_R}^2 = \vert M_N( A_N - \mu'\cot\beta') \vert$. It is clear that the lightest $\tilde{\nu}_{-}  \equiv \tilde{\nu}_{R_1}$ is the lightest sneutrino and can be even the LSP for a wide region of parameter space~\cite{Khalil:2011tb}, hence it can be stable and a viable candidate for DM. 

However, the interactions of sneutrino DM, $\tilde{\nu}_{R_1}$, are quite limited, therefore the annihilations of the $\tilde{\nu}_{R_1}$ are mainly given by  (when kinematically allowed) the four-point interaction: $ \tilde{\nu}_{R_1} \tilde{\nu}_{R_1 }\rightarrow h_i h_j$ and processes mediated by the CP-even Higgs sector $\tilde{\nu}_{ R_1} \tilde{\nu}_{R_1} \rightarrow  h_i \rightarrow h_i h_j~{\rm or} ~W^+ W^- $. The value of these annihilation
cross sections determine the relic abundance, which is given by 
\be 
\Omega_{\tilde{\nu}_{R_1}} h^2 = \frac{2.1 \times 10^{-27}~ {\rm cm}^3{\rm s}^{-1}}{\langle \sigma^{\rm ann}_{\tilde{\nu}_{R_1}} v \rangle} \left(\frac{x_F}{20}\right) \left(\frac{100}{g_*(T_F)}\right)^{\frac{1}{2}}
\ee  
where $\langle \sigma^{\rm ann}_{\tilde{\nu}_{R_1}} v \rangle$ is a thermal average for the total cross section of annihilation multiplied by the relative sneutrino velocity, $T_F$ is the freeze out temperature, $x_F =m_{\tilde{\nu}_{R_1}}/T_F \simeq {\cal O}(20)$ and $g_*(T_F ) \simeq  {\cal O}(100)$ is the number of degrees of freedom at freeze-out.
As emphazised in Ref.\cite{PhysRevD.76.041302} the suppressed annihilation cross sections lead to a very large relic abundance $\Omega_{\tilde{\nu}_{R_1}} h^2 $ (typically much larger than one), and for few points in the parameter space it can be within the $2\sigma$ allowed region by the Planck collaboration: $0.09 < \Omega h^2 < 0.14$.

Before closing this section, let us recall that the lightest $(B-L)$ Higgs is affected by the trilinear coupling,
\bea
{m}^2_{h'} &=& \frac{1}{2} \Big[ ( m^2_{A'} + M_{{Z'}}^2 ) - \sqrt{ ( m^2_{A'} + M_{{Z'}}^2 )^2 - 4 m^2_{A'} M_{{Z'}}^2 \cos^2 2\beta' }\;\Big].\nonumber
\eea
where $M_{A'}^2$ is the mass of the $(B-L)$ CP-odd Higgs,
\begin{equation}
M_{A'} = \frac{2B_{\mu '}}{\sin 2\beta'},
\end{equation}
and $B_{\mu '}$ is determined by the $B-L$ minimisation condition.
If $\cos^2{{2}\beta'} \ll 1$, one finds that the lightest $B-L$ neutral Higgs is given by %
\be%
{m}_{h'}\; {\simeq}\; \left(\frac{m^2_{A'} M_{{Z'}}^2 \cos^2 2\beta'}{{m^2_{A'}+M_{{Z'}}^2}}\right)^{\frac{1}{2}} \simeq {\cal O}(100~ {\rm GeV}).%
\ee%

As avocated above, in the BLSSM it is quite natural to have two light CP-even Higgs bosons $h$ and $h'$ with mass $125$ GeV and ${\cal O}(28)$ GeV, respectively. The CP-even neutral Higgs mass matrix can be diagonalised by a unitary transformation:
\be 
{\Gamma}~ M^2 ~ \Gamma^\dag = {\rm diag}\{m_h^2, m_H^2, m_{h'}^2, m_{H'}^2\}. 
\ee
The lightest $h'$ can be written in terms of  gauge eigenstates as 
\be 
h' = \Gamma_{31} ~\sigma_1 + \Gamma_{32} ~\sigma_2 + \Gamma_{33}~ \sigma'_1  +\Gamma_{34}~ \sigma'_2,
\ee 
where $\sigma_{1,2}$ are the real components of $H_{1,2}$ and $\sigma'_{1,2}$ are the real components of $\chi_{1,2}$, respectively. As $h'$ is essentially obtained from $B-L$ sector, the coupling $\Gamma_{33,34} \gg \Gamma_{31,32}$. In this regard, the $h'$ couplings with the ${W^{+}}{W^-}$ and $Z{Z}$ gauge bosons are given by:
\bea%
{\text g}_{W^\pm} &=& g_{{2}} M_W \left(\Gamma_{{32}} \sin \beta + \Gamma_{{31}} \cos\beta\right),\n
{\text g}_{ZZ} &\simeq& g_z\, M_Z  \left(\Gamma_{32} \sin{\beta}+ \Gamma_{31} \cos{\beta}\right),
\eea
%%%%%%%%%%%%%%%%%%%%%%%%%%%%%%%%%%%%%%
\section{ $\tilde{\nu}_{ R_1}$  Bound state and Sommerfeld Effect} 

\begin{figure}[h!]
\begin{center}
\includegraphics[width=0.3\textwidth]{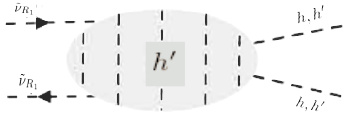}
\end{center}
\vskip -0.5cm
\caption{$\tilde{\nu}_{R_1}$ annihilation to $hh$, $h'h'$ and $W^+W^-$ through the bound state mediated by $h'$ Higgs boson before annihilation.}
\label{fig:Bh}
\end{figure}

As both $h'$ and $\tilde{\nu}_{ R_1}$ belong to the B-L sector, the have a reasonably large strength of their interaction. Thus, a bound state can be formed for $\tilde{\nu}_{ R_1}$ by mediating the scalar boson $h'$ as shown in Fig. \ref{fig:Bh}.  

%%%%%%%
%%%%%%

As is well known, scalar particle exchange mediates a Yukawa potential. In this regard, the sneutrino bound state mediated by $h'$ Higgs boson, yields a Yukawa potential with range $1/m_{h'}$, is only possible if  
\be
\frac{\alpha_\nu ~m_{\tilde{\nu}_{ R_1}}}{m_{h'}} > 1.68,
\ee
where $\alpha_\nu$ is the coupling of $\tilde{\nu}_{ R_1}\tilde{\nu}_{ R_1}h'$ \cite{Kang:2016wqi}.  Thus, due the large $m_h'$, the sneutrino bound state can be achieved only for heavy sneutrino or/and  a quite strong coupling.   

The internal motion of the bound state of $\tilde{\nu}_{ R_1}$ is non-relativistic that can be described by Schrodinger equation 
\be 
\left( - \frac{\bigtriangledown^2}{2 \mu_r} + V(r) \right) \psi(r) = E \psi(r), 
\ee
where $\mu_r = m_{\tilde{\nu}_{ R_1}}/2$ is the reduced mass and the $V(r)$ is the corresponding Yukawa potential, obtained from $h'$-exchange and given by 
\begin{equation}
V(r)=- \alpha_\nu \frac{e^{-m_{h'} r}}{r}, 
\end{equation}
where $\alpha_\nu= (Y)^2/16 \pi m^2_{\tilde{\upsilon}_1} $. The details of the coupling $Y$ will be discussed in the next section. 

The effect of non-relativistic Yukawa potential, due to the large distance force between the incoming sneutrinos, on the annihilation cross section can be represented by the ratio between the distorted wave function to the unperturbed one at the origin (annihilation point). Therefore, the annihilation cross section with bound state can be written in terms of the usual annihilation cross section as follows: \cite{MarchRussell:2008yu,Iengo:2009ni, Cassel:2009wt}
\be 
\sigma = S \sigma_0,
\ee
where $S$ is known as Sommerfeld effect, which is given for the s-wave case by \cite{ Plehn:2017fdg, Iengo:2009ni, MarchRussell:2008yu}
\be S = \frac{\vert \psi(0) \vert^2}{\vert \psi^0(0)\vert^2} =\vert \psi(0) \vert^2 .
\ee 
Here we assume that $\vert \psi^0(0) \vert^2 =1$.  Since the exact analytical solution of Schrodinger equation with Yukawa potential is not available, it is common to use instead an approximate solution that can be obtained if one considered a modified version of $V(r)$ as follows:
\be 
V_H(r)=-\alpha m_* \frac{e^{-m_*~r}}{1-e^{-m_*~r}},
\ee
where $m_* \simeq \frac{\pi^2}{6} m_{h'} $ \cite{Oncala:2018bvl}. This type of potential is called Hulthen potential. Now, it is possible to obtain analytical solutions for the $\ell = 0$ modes of the wavefunctions. It turns out that the s-wave Sommerfeld enhancement is given by \cite{Oncala:2018bvl,Liu:2013vha,Plehn:2017fdg}
\be 
S =(\frac{2 \pi \alpha}{\upsilon}) \frac{sinh( \frac{6 m_{\tilde{\upsilon}_{1}} \nu_{rel}}{\pi m_{h'}})}{cosh(\frac{6 m_{\tilde{\upsilon}_{1}} \nu_{rel}}{\pi m_{h'}})-cos[\sqrt{\frac{24 m_{\tilde{\upsilon}_{1}}  \alpha }{ m_{h'}}-\frac{36 m^2_{\tilde{\upsilon}_{1}} \nu^2_{rel}}{\pi^2 m^2_{h'}}]}},
\ee
where $v_{rel}$ is the relative velocity between the $\tilde{\nu}_{R_1}$ DM particles. Since $v_{rel} \ll 1$, the Sommerfeld effect can be approximate to 
\be 
S = \frac{2\pi \alpha_\nu}{ v_{rel}}  \frac{6 m_{\tilde{\nu}_{R_1}} v_{rel}}{1- \cos\left( \sqrt{\frac{24 m_{\tilde{\nu}_{R_1}} \alpha_\nu}{m_{h'}}}\right)},
\ee

%%%%%%
\begin{figure}[h!]
\begin{center}
\includegraphics[width=0.5\textwidth]{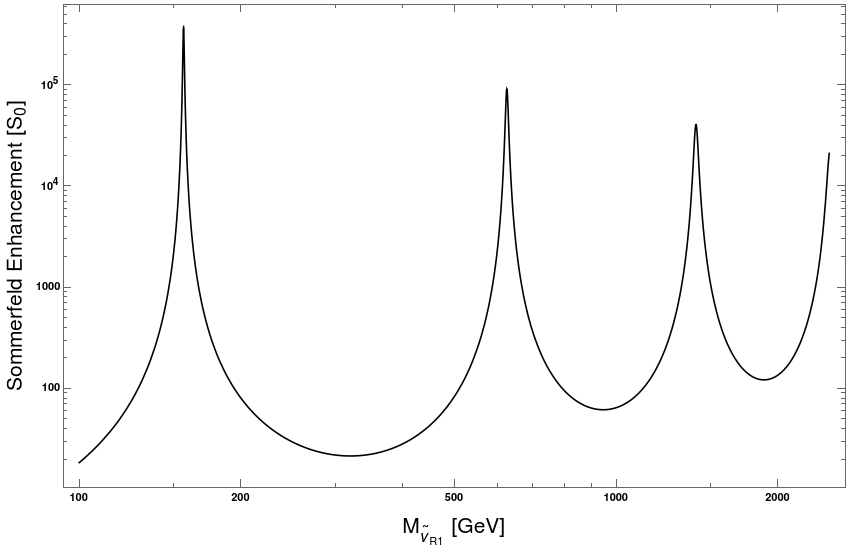}
\end{center}
\vskip -0.5cm
\caption{Sommerfeld enhancement for a Hulthen potential as a function of $m_{\tilde{\upsilon}_{R_1}}$, with $ m_{h'}=28$ Gev and a coupling strength of $\alpha_\nu= 0.3$.}
\label{fig:sommerfeld}
\end{figure}
%%%%%

In Fig. \ref{fig:sommerfeld}, we display the Sommerfeld enhancement for a Hulthen (Yukawa) potential as a function of the $\tilde{\nu}_{R_1}$ and coupling $\alpha_\nu$ of order $0.3$. It is interesting to note that Sommerfeld effect could be huge and the peaks in the curve correspond to 
$\cos\left( \sqrt{\frac{24 m_{\tilde{\nu}_{R_1}} \alpha_\nu}{m_{h'}}}\right)=1$, {\it i.e.}, $\frac{24 m_{\tilde{\nu}_{R_1}} \alpha_\nu}{m_{h'}}=(2 n \pi)^2$ and $n$ is integer.

This enhancement of annihilation cross section will have a significant effect on computing the relic abundance, $\Omega h^2$ of $\tilde{\nu}_{R_1}$ DM, as it is inversely proportional to the thermal average of $(\sigma_{ann} v_{rel})$. In general, the usual DM annihilation cross section times the relative velocity can be written as $(\sigma v_{rel})_0=a+b v^2_{rel}$, where $a$ and $b$ are the 	$S$- and $P$-wave contributions, respectively. With Sommerfeld effect, the thermally averaged cross section of $S$-wave at a temperature $T$ or $x\equiv m_\chi/T$ can be written as \cite{Liu:2013vha}
\bea
\langle \sigma v_{rel} \rangle&=&a \langle S_0(v_{rel})\rangle (x) \\
&=& \frac{(\sigma \upsilon)_0 ~ x^{3/2}}{2\sqrt{\pi}} \int_{0}^{\infty}  ~S_0(\upsilon_{rel})~ e^{- \frac{x \upsilon_{rel}^2}{4}} \upsilon^2_{rel}~d\upsilon_{rel}   
\label{S_0}\nonumber
\eea
%

%%%%%%%%%%%%%%%%%%%%%%%%%%%%%
\section{Revisiting Right-handed Sneutrinos DM}

In the BLSSM, the relevant interaction terms of lightest right-handed sneutrino are given by the following Lagrangian:
\bea
\mathcal{L} = Y (\tilde{\upsilon}_{ R_1})^2 h'+\lambda_{4}  (\tilde{\upsilon}_{R_1})^2 h'^2 + \lambda_{2} h'h^2 +\textsl{g}_{W^\pm} h' W^+ W^-  +\lambda_{3} h'h'h' + \textsl{g}_{ZZ} h' Z Z  ,
\eea
where the above couplings are defined as follows:

\begin{align}
\textsl{g}_{W^\pm } &\simeq g_{{2}} M_W \left(\Gamma_{{32}} \sin \beta + \Gamma_{{31}} \cos\beta\right),~~~ \textsl{g}_{ZZ} \simeq g_z\, M_Z  \left(\Gamma_{32} \sin{\beta}+ \Gamma_{31} \cos{\beta}\right),\\
Y&\simeq (\Gamma^{R}_{1 4})^2  \Big[\frac{g_{BL}^{2}}{2} \Big( v'_1 \Gamma_{{3 3}} - v'_2 \Gamma_{{3 4}} \Big) + \sqrt{2} \Big( \Gamma_{{3 4}} \mu' Y_N - \Gamma_{{3 3}} T_N \Big) -4v'_1 \Gamma_{{3 3}} Y_N^2 \Big], \\ 
\lambda_{4}&\simeq (\Gamma^{R}_{1 4})^2  \Big[\frac{g_{BL}^{2}}{2} \Big(\Gamma_{{3 3}}^2 -\Gamma_{{3 4}}^2 \Big) + \frac{g_{BL} \tilde{g}}{4} \Big( \Gamma_{{3 1}}^2 - \Gamma_{{3 2}}^2 \Big) - 4 \Gamma_{{3 3}}^2 Y_N^2 \Big], \\
\lambda_{3}&\simeq g_{BL}^{2} \Big[ v'_1\Big(-3\Gamma_{{3 3}}^{3} + 3 \Gamma_{{3 3}} \Gamma_{{3 4}}^{2}\Big) 
+v'_2 \Big(3 \Gamma_{{3 3}}^2 \Gamma_{{3 4}} -3 \Gamma_{{3 4}}^3 \Big) \Big],\\
\lambda_{2}&\simeq g_{BL}^{2} \Big[ v'_1\Big(-3\Gamma_{{3 3}}^{2} \Gamma_{{1 3}} + \Gamma_{{1 3}} \Gamma_{{3 4}}^{2} +2 \Gamma_{{3 4}} \Gamma_{{3 3}} \Gamma_{{1 3}}\Big) 
+v'_2 \Big(  \Gamma_{{3 3}}^2 \Gamma_{{1 4}} +2\Gamma_{{3 3}} \Gamma_{{3 4}} \Gamma_{{1 3}} -3 \Gamma_{{3 4}}^2  \Gamma_{{1 4}} \Big) \Big]
\end{align}
Here $\Gamma$ and  $\Gamma^{({\rm R})}$ are the matrices that diagonalize the CP-even Higgs mass matrix and right-handed sneutrino mass matrix respectively.  There are four Higgs VEVs, corresponding to the MSSM $H_u$ and $H_d$ doublets and the $B-L$ $\chi_1$ and $\chi_2$ singlets, written as $\left( v_u,~v_d,~v'_1,~v'_2 \right)$, respectively.  Finally, $T_{N}$ is the trilinear couplings, which is defined as $Y_N A_N$. 

\begin{figure}[h!]
\begin{center}
\includegraphics[width=0.6\textwidth]{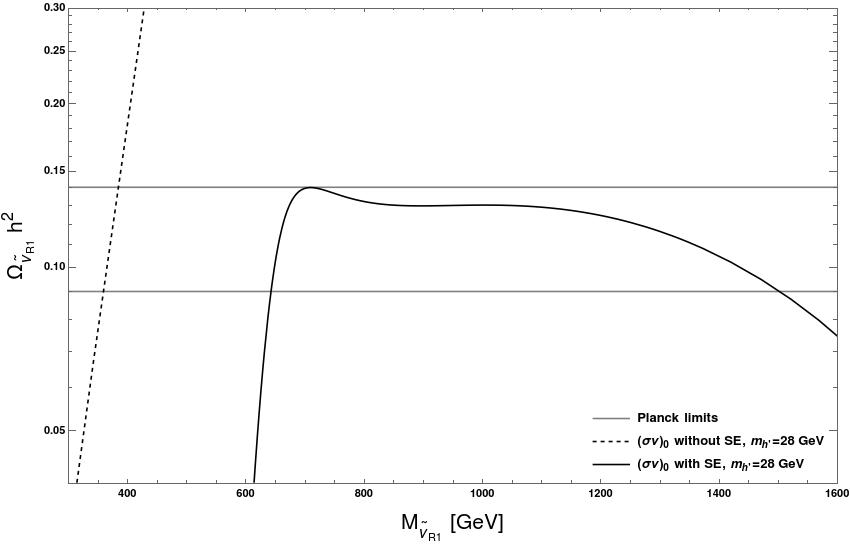}
\end{center}
\vskip -0.5cm
\caption{(Dashed plot) Relic density of the dark matter mass $\tilde{\nu}_{R_{1}}$ without Sommerfeld enhancement, (Solid plot) with Sommerfeld enhancement for $m_{h'}=28$ GeV, $Y \sim 2 m_{\tilde{\nu}_{R_{1}}} \gsim \sqrt{16\times 1.68 \times \pi m_{h '} m_{\tilde{\nu}_{R_{1}}}}  $. The horizontal lines correspond to the Planck limits for the relic abundance.}
\label{300SE}
\label{fig:sommerfeld}
\end{figure}

The numerical values of the above parameters are computed by SARAH and SPheno programs \cite{Staub:2013tta, Porod:2003um}. In Fig. \ref{300SE}, we present the relic abundance of $\tilde{\nu}_{R_1}$ DM with and without Sommerfeld enhancement as function of $\tilde{\nu}_{R_1}$ mass for the following benchmark point: $g_{BL}=0.55,~\tilde{g}=-0.138,\tan \beta =40,~\tan \beta' =1.19, ~v'\simeq 7.2 ~{\rm TeV},~\mu'\simeq 12.5~{\rm TeV},~T_N \simeq -12.5~{\rm TeV},Y_N \simeq {\cal O}(1), ~\Gamma_{31}\simeq -0.002,~\Gamma_{32}\simeq-0.006,\Gamma_{33}\simeq-0.774,~\Gamma_{34}\simeq 0.66,~\Gamma_{11}\simeq -0.03,~\Gamma_{12}\simeq -0.99 ~\Gamma_{13}\simeq 0.005,\Gamma_{14} \simeq 0.002$,
which leads to $Y \simeq 2.5~ {\rm TeV},~\lambda_3 \simeq 179~ {\rm GeV},~ \lambda_{4}\simeq-4.5 ,~\lambda_{2} \simeq 0.16~ {\rm GeV},~\textsl{g}_{w^\pm}\simeq -0.313,\textsl{g}_{ZZ}\simeq -0.405,$
%%%%%%%%% 
%%%%%%%%%

From this figure , it is clear that the relic abundance can only be consistent with the observational limits without the Sommerfeld effect for a very special values of right-handed sneutrino mass that render certain fine tuning. However, with Sommerfeld effect these constraints are met for a wider region of $m_{\tilde{\nu}_{R 1}}$ and right-handed sneutrino with mass of order one TeV remains a viable DM.

\section{Conclusions}
In this paper we have analyzed the possibility that the lightest right-handed sneutrino in the BLSSM can form a bound state through the exchange of the lightest CP-even Higgs boson, $h'$, which is associated with  $U(1)_{B-L}$ symmetry breaking. This Higgs boson can be as light as $28$ GeV with a large coupling to the right-handed sneutrino, hence it naturally  mediates the lightest right-handed sneutrino bound state.  We have shown that the resultant Sommerfeld Effect can be quite large and would provide a major increase in the cross section of annihilation. We emphasized that in general and without bound state effect the relic abundance of right-handed sneutrino is quite large and exceed the observational limit for most of the parameter space of the BLSSM model, as it is inversely proportional to a very small annihilation cross section. We have proven that with Sommerfeld effect, we could have the right-handed sneutrino as a viable DM candidate, with relic abundance within the experimental limits: $\Omega h^2 = 0.12 \pm 0.002$, for masses between 600 GeV and 1.5 TeV. 
\section*{Acknowledgments}
We would like to thank N. Brambilla, A.Vairo and G.Qerimi for useful discussions.

\nocite{*}
\bibliographystyle{unsrt} 
\bibliography{Refs}
\end{document}